\documentclass[a4paper,12pt]{revtex4}

\usepackage{latexsym}
\usepackage{graphics}
\usepackage{color}
\usepackage[latin1]{inputenc}
\usepackage{amssymb}
\usepackage{amsmath}
\usepackage{amsfonts}
\usepackage[dvips]{graphicx}

\usepackage{pdfpages}
\usepackage{epstopdf}

\begin{document}

\title{Differential phase extraction in an atom gradiometer}

\author{F. Pereira Dos Santos}

\email[]{franck.pereira@obspm.fr}

\affiliation{LNE-SYRTE, Observatoire de Paris, LNE, CNRS, UPMC, 61 avenue de l'Observatoire, 75014 Paris, France}

\begin{abstract}
We present here a method for the extraction of the differential phase of an atom gradiometer that exploits the correlation of the vibration signal measured by an auxiliary classical sensor, such as a seismometer or an accelerometer. We show that sensitivities close to the quantum projection noise limit can be reached, even when the vibration noise induces phase fluctuations larger than $2\pi$. This method doesn't require the correlation between the atomic and classical signals to be perfect and allows for an exact determination of the differential phase, with no bias. It can also be applied to other configurations of differential interferometers, such as for instance gyrometers, conjugate interferometers for the measurement of the fine structure constant, or differential accelerometers for tests of the equivalence principle or detection of gravitational waves.

\end{abstract}

\pacs{37.25.+k, 06.30.Gv, 04.80.-y, 03.75.Dg}

\maketitle

\section{Introduction}

Atom interferometers have demonstrated performances comparable or better than state-of-the-art classical instruments, both in terms of sensitivity and accuracy, and find applications in various fields, from fundamental physics to geophysics and navigation. In particular, differential atom interferometers allow for an improved determination of the quantity to be measured, as their mode of operation rejects common mode noise sources and systematic effects. They are used for the measurement of gravity gradients \cite{McGuirk2002,Sorrentino2014} and rotation rates \cite{Gustavson1997,Canuel2006,Berg2015}, for the determination of $G$ \cite{Fixler2007,Rosi2014} and the ratio $h/m$ \cite{Bouchendira2011,Lan2013}, and for tests of the universality of free fall \cite{Dimopoulos2007,Bonnin2013,Aguilera2014} and are expected to find applications in the detection of gravitational waves \cite{Delva2006,Tino2007,Dimopoulos2008,Graham2013}. In most cases, the fluctuations of the common phase, in general dominated by the effect of parasitic vibrations, wash out completely the visibility of single interferometers, but the correlation between the two output signals can be used to recover the information on the difference between the two interferometer phases. Various techniques have been demonstrated to extract the differential phase. A method based on ellipse fits was first used in \cite{Foster2002}, but was shown to introduce significant biases, that are errors in the determination of this differential phase. Methods based on Bayesian estimators are more accurate \cite{Stockton2007,Varoquaux2009,Chen2014}, but require an a priori model of the interferometer phase noise. In a recent study, it was shown that the use of a simultaneous third measurement allows for the retrieval of the differential phase(s) with negligible bias(es) using a robust three dimensional fit \cite{Rosi2015}.

In this paper, we present an alternative method, that uses the correlation of the interferometer phases with an estimate of the vibration phase provided by an auxiliary classical sensor. It allows to recover the differential phase of interferometers operated in a gradiometer configuration, without any biasing or an a priori knowledge of the noise distribution. We show in particular that sensitivities close the quantum projection noise limit can be obtained.

\section{Principle of the method}

The output signals of the interferometers are given by the transition probabilities $P_1$ and $P_2$:
\begin{align*} 
P_1=A_1+\frac{C_1}{2}\textrm{cos}(\phi_m+\phi_1+\phi_{1l} )\\
P_2=A_2+\frac{C_2}{2}\textrm{cos}(\phi_m+\phi_2+\phi_{2l})\\
\end{align*}
where 
$\phi_m$ is the phase shift induced by the vibrations (of the mirror used to retroreflect the lasers of the interferometer), $\phi_i$ are the two interferometer phase shifts (with contributions from inertial effects, such as gravity acceleration, and non-inertial effects, such as lights shifts) and $\phi_{il}$ are controlled phases shifts applied onto the phase of the interferometers. $\phi_{il}$ can be common to the two interferometers (such as when applying to the laser phase difference either a frequency chirp for Doppler shift compensation or a phase jump), or not. In particular, such differential phase shifts can easily be set in the configuration of a gradiometer, taking advantage of the spatial separation between the interferometers. This was realized for instance by applying a magnetic pulse at one of the two clouds \cite{McGuirk2002,Duan2014}, or by inducing a common frequency jump onto the interferometer lasers \cite{Biedermann2014}, the latter technique providing in principle a better control of the applied differential phase.

Without loss of generallity, we will take $\phi_1+\phi_{1l}=0$. This can easily be realized, for instance by applying a phase jump before the last pulse of the interferometer. This leads to 
\begin{align*} 
P_1=A_1+\frac{C_1}{2}\textrm{cos}(\phi_m) \\
P_2=A_2+\frac{C_2}{2}\textrm{cos}(\phi_m+\phi_d) \\
\end{align*}
where $\phi_d=\phi_2-\phi_1+\phi_{2l}-\phi_{1l}$ is the differential phase between the two interferometers. 

An estimate $\phi_s$ of the phase shift induced by the vibration of the mirror $\phi_m$ can be determined thanks to the measurement of an auxiliary sensor, such as a seismometer \cite{LeGouet2008} or a mechanical accelerometer \cite{Geiger2011,Lautier2014}, or with an additional interferometer as in \cite{Sorrentino2012}. When using a classical sensor, a rigid link with the mirror is required for an optimal correlation between $\phi_m$ and $\phi_s$, and some signal processing can be used to flatten the transfer function of the sensor \cite{LeGouet2008}. 

This correlation can be exploited to determine the phase of the interferometer, even in the presence of large vibration noise that washes out the fringes at the output of the interferometer. One can exploit for instance the fringe fitting technique described in \cite{Merlet2009}, that is based on the recovery of interferometer fringes when plotting the transition probability versus $\phi_s$. Vibration noise provides there a random sampling of the interferometer phase that allows the fringe pattern to be scanned.

We show here that this technique can be extended to the case of a differential interferometer, by reconstructing the fringes for both interferometers independently, and fitting them as a function of $\phi_s$. Though the dispersion of the phases of the two interferometers, as determined by the fits, is linked to the quality of the correlation between real and measured mirror vibrations (and might thus exhibit relatively large fluctuations when this correlation is poor), the differential phase can potentially be much better determined, because the fluctuations of the fitted phases of the two interferometers are also correlated.

\section{Simulation}

To simulate the signals at the output of the interferometer, we start by randomly drawing the values of $\phi_m$ in a gaussian distribution with a standard deviation of $\sigma_{\phi_m}$. In order to account for an imperfect correlation between the mirror motion and the sensor signal, we randomly draw the difference between the corresponding phases $\delta\phi=\phi_m-\phi_s$, with a standard deviation of $\sigma_{\delta\phi}$. The two transition probabilities are then calculated, with $A_1=A_2=0.5$ and $C_1=C_2=1$. We then account for the influence of detection noise, that we take as quantum projection noise limited with a number of detected atoms of $N_{at}=10^6$ for each interferometer. We thus add to $P_i$ noise contributions $\delta_{P_i}$, randomly drawn in gaussian distributions with standard deviations $\sigma_{P_i}=\sqrt{P_i(1-P_i)/N_{at}}$. 

\section{Results}

\begin{figure}[ht]
        \centering
      \includegraphics[width=12 cm]{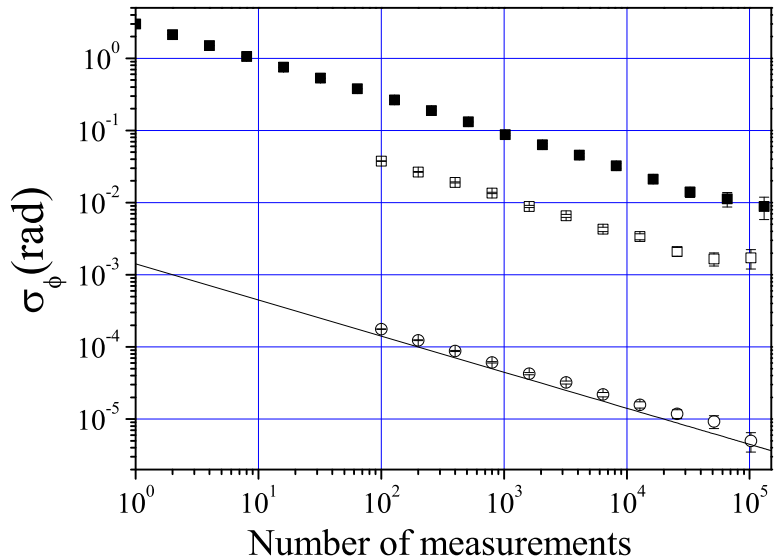}
\caption{Allan standard deviation of the differential phase (open circles), compared to Allan standard deviations of the phase of a single interferometer (open squares) and of the vibration noise (black squares). The line represents the quantum projection noise limit. Parameters are $\sigma_{\phi_m}=3$ rad, $\sigma_{\delta\phi}=0.3$ rad and $\phi_d=0$.}
        \label{fig:var}
   \end{figure}

First, we generate $10^6$ pairs $(P_1,P_2)$ with $\sigma_{\phi_m}=3$ rad, $\sigma_{\delta\phi}=0.3$ rad and $\phi_d=0$. Such a level of vibration noise and degree of correlation between mirror vibrations and sensor signal corresponds to the case studied in \cite{Merlet2009}, where an atom gravimeter operating in a urban environment was directly put on the ground and the classical sensor was a seismometer. We then group the data by sets of 100 trials and perform $10^4$ consecutive fits of $P_i$ versus $\phi_s$, from which extract $10^4$ values of the fitted phases $\phi_{i,f}$, and of their difference $\phi_{d,f}=\phi_{2,f}-\phi_{1,f}$. Figure \ref{fig:var} displays the Allan standard deviation of the difference of the fitted phases $\sigma_{\phi_{d,f}}$ (open circles), compared with the fitted phase of one of the interferometers (open squares) and with the vibration phase noise (full squares). While the correlation with the data of the classical sensor allows to gain a factor of about 8 on the sensitivity of a single interferometer, the gain on the differential phase noise is much larger, of about 1700. The thin line represents the quantum projection noise limit of $\sigma_{QPN}=\sqrt{2}/\sqrt{N_{at}}$, and only lies a factor 1.25 below. 

\begin{figure}[ht]
        \centering
      \includegraphics[width=16 cm]{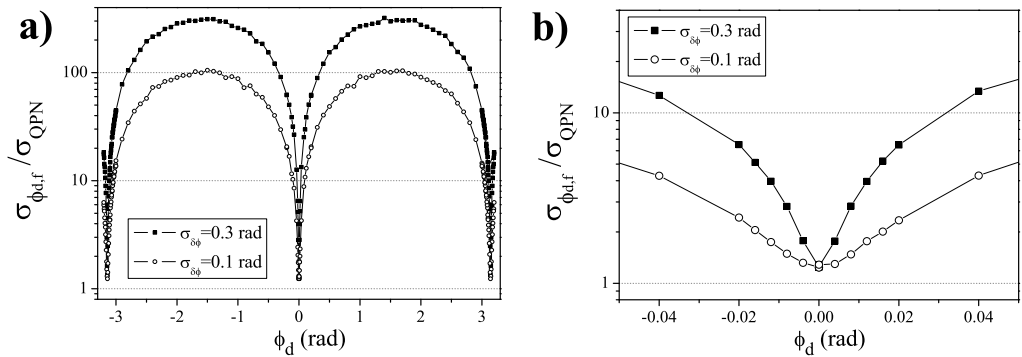}
\caption{Sensitivity in the differential phase extraction, normalized to the quantum projection noise limit as a function of the value of the differential phase. A zoom on the curve for small values of the differential phase, below 50 mrad, is displayed on the right.}
        \label{fig:vsPhid}
   \end{figure}

We then investigate the efficiency of this rejection as a function of the value of the differential phase, by plotting the ratio $\sigma_{\phi_{d,f}}/\sigma_{QPN}$ as a function of the differential phase $\phi_d$, for a fixed value of $\sigma_{\phi_m}=3$ rad, and different amplitudes of $\sigma_{\delta\phi}$. The results are displayed on figure \ref{fig:vsPhid}. We observe an optimal phase sensitivity when $\phi_d=0[\pi]$, close to the quantum projection noise limit. We observe a rapid degradation as the value of $\phi_d$ starts deviating from these values, that increases with $\sigma_{\delta\phi}$. In particular, the correlation between the two fitted phases is lost when operating the two interferometers in quadrature. There, the sensitivity is minimal and comparable to the sensitivity one would obtain with independent uncorrelated interferometers. 

\begin{figure}[ht]
        \centering
      \includegraphics[width=12 cm]{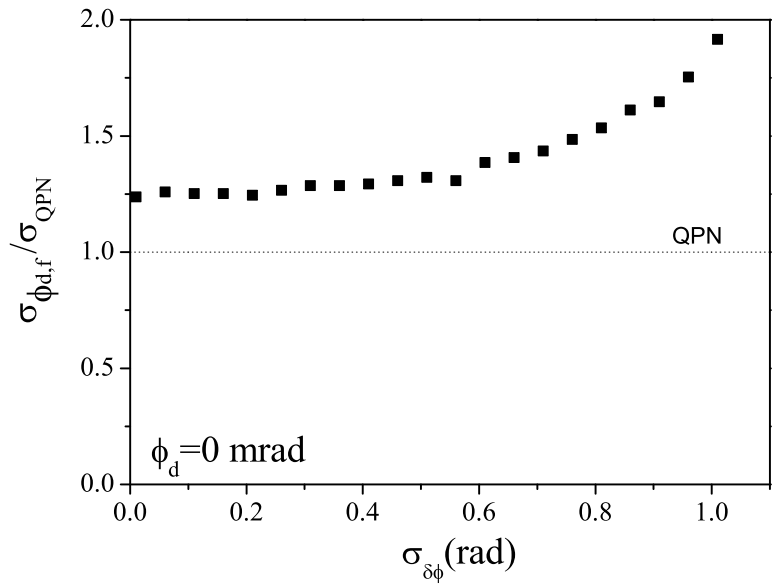}
\caption{Sensitivity in the differential phase extraction, normalized to the quantum projection noise limit, as a function of the standard deviation of $\delta\phi$, the difference between real and measured mirror vibrations. The differential phase is zero.}
        \label{fig:vsDeltaPhi}
   \end{figure}

In a second series of simulations, we vary the amplitude of $\sigma_{\delta\phi}$, keeping $\phi_{d}=0$ and $\sigma_{\phi_m}=3$ rad. The results displayed on figure \ref{fig:vsDeltaPhi} show that the sensitivity in the differential phase remains close to the QPN limit, within a factor of 2, over a large range of $\sigma_{\delta\phi}$ of 1 radian. This shows that the retrieval of the differential phase is still possible, even when the correlation between the classical signal and the mirror motion is far from being perfect and the level of uncorrelated noise is high. Degradation of the correlation can arise from non linearity of the sensor in the frequency range where the vibration noise is important (from 0.1 to 100 Hz), or from the intrinsic noise of the classical sensor. In the example above, a level of intrinsic noise of the classical sensor lower than the level of vibration noise by a factor of only 3 would still allow reaching a sensitivity as good as twice the quantum projection noise limit. This shows that, depending on the parameters, mid class sensors, which are more compact and less expensive, can in practice still be highly beneficial.

\begin{figure}[ht]
        \centering
      \includegraphics[width=12 cm]{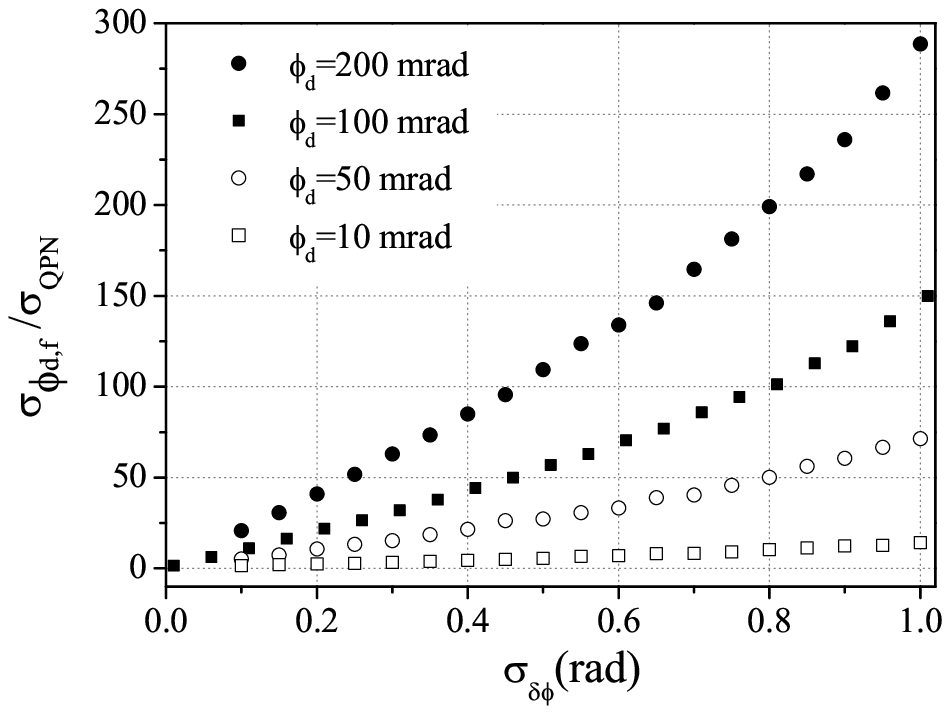}
\caption{Sensitivity in the differential phase extraction, normalized to the quantum projection noise limit, as a function of the standard deviation of $\delta\phi$, the difference between real and measured mirror vibrations. The differential phase is varied between 10 mrad and 200 mrad.}
        \label{fig:vsDeltaPhi2}
   \end{figure}

We then repeat these simulations for increasing values of $\phi_d$, ranging from 10 mrad to 200 mrad. Figure \ref{fig:vsDeltaPhi2} shows the results, that illustrate that the trend in the degradation of the sensitivity versus $\delta\phi$ increases with increasing values of $\phi_d$, which was already observed in figure \ref{fig:vsPhid}. 

\begin{figure}[ht]
        \centering
      \includegraphics[width=12 cm]{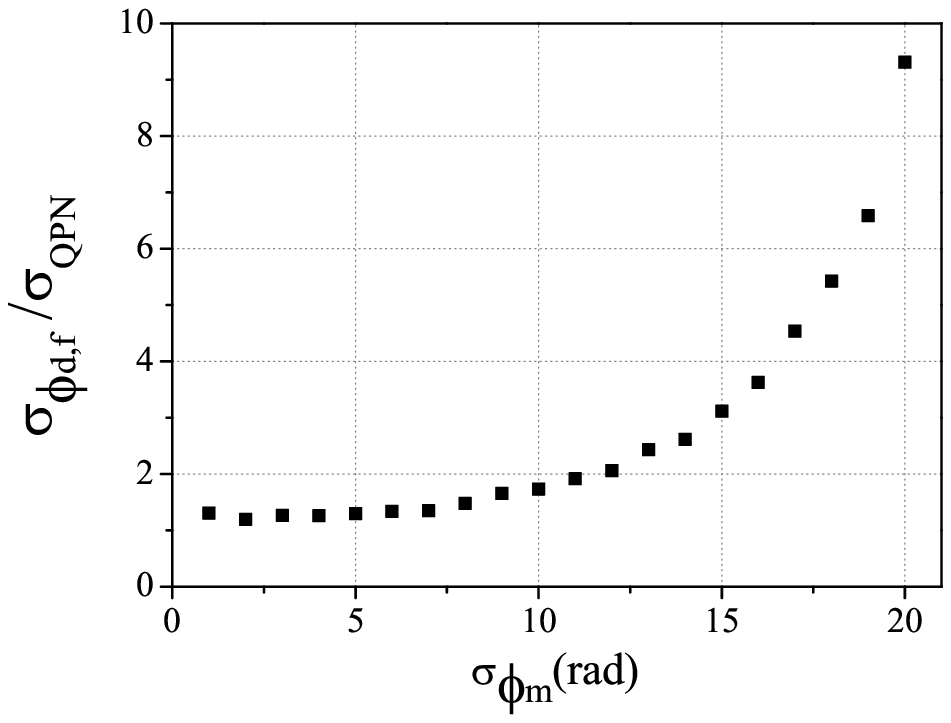}
\caption{Sensitivity in the differential phase extraction, normalized to the quantum projection noise limit, as a function of the mirror vibration noise $\sigma_{\phi_m}$. The ratio between $\sigma_{\phi_m}$ and $\sigma_{\delta\phi}$ is kept constant, equal to 10. The differential phase is null.}
        \label{fig:vsPhim}
   \end{figure}

Finally, we vary the amplitude of  $\sigma_{\phi_m}$ keeping the quality of the correlation the same, which means that we keep the ratio $\sigma_{\phi_m}/\sigma_{\delta\phi}$ constant, equal to 10. Figure \ref{fig:vsPhim} shows an increase by less than a factor 2 with $\sigma_{\phi_m}$ below an amplitude of 10 rad. Note that for such a value, $\sigma_{\delta\phi}=1$ rad which corresponds to a situation where the fringes obtained when plotting $P_i$ versus $\phi_m$ are hardly visible. Even in that case, the sensitivity is only twice the QPN limit, and the rejection efficiency of vibration noise is as large as a factor 4000. For larger vibration noise, the degradation of the sensitivity gets significant. Also, the number of points to be fitted needs to be increased for the fit to converge to the correct value. We obtain stable fits up $\sigma_{\delta\phi}=20$ rad with a number of points to fit of 1000. 

Of importance, we checked that, whatever the noise parameters we choose, the extraction of the differential phase is free from any biasing: the mean value of the fitted differential phase is always found equal to the differential phase. This will hold in a practical implementation in particular when the mean value of $\phi_s$ and $\phi_m$ are equal (and for instance equal to zero in the absence of a long term motion of the mirror). This equality is verified if using a low noise seismometer \cite{LeGouet2008,Merlet2009}, whose output signal is a velocity signal, but not necessarily if using a DC accelerometer, such as in \cite{Lautier2014}. In the latter case, a non zero DC value leads to an offset in the fitted phase, that can exhibit long term drifts. Nevertheless, this offset being common to both interferometers, it will cancel out when calculating the differential phase. Finally, we checked that the results presented here do not depend on the values of the offsets $A_1$ and $A_2$ on the transition probabilities, nor on the values of $C_1$ and $C_2$ (if accounting for the corresponding degradation of the QPN limited sensitivity).  

\section{Conclusion}

We presented here a technique for extracting the differential phase in an atom gradiometer in the presence of large vibration noise, that washes out completely the visibility of the fringes of the individual interferometers. It relies on the exploitation of the auxiliary signal provided by a classical sensor, that measures the motion of the mirror that retroreflects the interferometer lasers. We show that sensitivities close to the QPN limit can be reached, provided that the value of the differential phase modulo $\pi$ is zero. This corresponds to in-phase or in counter-phase operation of the two interferometers, which can be set either by changing the scale factor of the interferometers, or by applying a perturbation that is not common mode to the two interferometers.
The technique can also be used for other differential interferometers, such as gyrometers based on counterpropagating atomic sources \cite{Canuel2006,Berg2015}, or conjugate interferometers for the determination of the ratio $h/m$ \cite{Chiow2009}.

This technique is also of interest for differential accelerometers that use two different atomic species in order to perform tests of the equivalence principle. In that case though, the accelerometer scale factors may differ due to the difference in the effective wave-vectors of the interferometer lasers. Vibration phases, even if proportional, are thus different for each interferometer \cite{Chen2014}. As an alternative, the scale factor can be matched by making the interferometer pulse sequence slightly different for the two accelerometers \cite{Varoquaux2009}. Though the response of the two interferometers to DC acceleration are then identical, the difference in the pulse sequence makes the transfer function not identical for the two accelerometers and thus their response to vibration noise is different \cite{Barrett2015}. Even though the correlation between the fitted phases is then reduced, a recent study \cite{Barrett2015} demonstrates that the present technique compares favorably with other methods, and allows for a more robust extraction of the differential phase, free of any bias, and with a potentially excellent sensitivity.

\section{Acknowlegments}
The author thanks S. Merlet and A. Landragin for useful discussions and careful reading of the manuscript. This work was supported by the European Space Agency through the Contract No. 4000112677/14/NL/MP.


\begin{thebibliography}{0}
\expandafter\ifx\csname natexlab\endcsname\relax\def\natexlab#1{#1}\fi
\expandafter\ifx\csname bibnamefont\endcsname\relax
  \def\bibnamefont#1{#1}\fi
\expandafter\ifx\csname bibfnamefont\endcsname\relax
  \def\bibfnamefont#1{#1}\fi
\expandafter\ifx\csname citenamefont\endcsname\relax
  \def\citenamefont#1{#1}\fi
\expandafter\ifx\csname url\endcsname\relax
  \def\url#1{\texttt{#1}}\fi
\expandafter\ifx\csname urlprefix\endcsname\relax\def\urlprefix{URL }\fi
\providecommand{\bibinfo}[2]{#2}
\providecommand{\eprint}[2][]{\url{#2}}

\end{thebibliography}


\begin{thebibliography}{30}


\bibitem{McGuirk2002}  J. M. McGuirk, G. T. Foster, J. B. Fixler, M. J. Snadden, M. A. Kasevich, Phys. Rev. A {\bf 65}, 033608 (2002)

\bibitem{Sorrentino2014} F. Sorrentino, Q. Bodart, L. Cacciapuoti, Y.-H. Lien, M. Prevedelli, G. Rosi, L. Salvi, and G. M. Tino, Phys. Rev. A {\bf 89}, 023607 (2014)

\bibitem{Gustavson1997} T. L. Gustavson, P. Bouyer, and M. A. Kasevich, Phys. Rev. Lett. {\bf 78}, 2046 (1997)

\bibitem{Canuel2006} B. Canuel, F. Leduc, D. Holleville, A. Gauguet, J. Fils, A. Virdis, A. Clairon, N. Dimarcq, Ch. J. Bord\'e, A. Landragin, P. Bouyer, Phys. Rev. Lett. {\bf 97}, 010402, (2006) 

\bibitem{Berg2015} P. Berg, S. Abend, G. Tackmann, C. Schubert, E. Giese, W.P. Schleich, F.A. Narducci, W. Ertmer, and E.M. Rasel, Phys. Rev. Lett. {\bf 114}, 063002 (2015)

\bibitem{Fixler2007} J. B. Fixler, G. T. Foster, J. M. McGuirk, and M. A. Kasevich, Science {\bf 315}, 74 (2007)

\bibitem{Rosi2014} G. Rosi, F. Sorrentino, L. Cacciapuoti, M. Prevedelli, G.M. Tino, Nature {\bf 510}, 518-521 (2014)

\bibitem{Bouchendira2011} R. Bouchendira, P. Clad\'e, S. Guellati-Kh\'elifa, F. Nez, and F. Biraben, Phys. Rev. Lett. {\bf 106}, 080801 (2011)
 
\bibitem{Lan2013} S. Y. Lan, P. C. Kuan, B. Estey, D. English, J. M. Brown, M. A. Hohensee, and H. M{\"u}ller, Science {\bf 339}, 554 (2013)
 
\bibitem{Dimopoulos2007} S. Dimopoulos, P. W. Graham, J. M. Hogan, and M. A. Kasevich, Phys. Rev. Lett. {\bf 98}, 111102 (2007)

\bibitem{Bonnin2013} A. Bonnin, N. Zhazam, Y. Bidel, A. Bresson, Phys. Rev. A {\bf 88}, 043615 (2013)

\bibitem{Aguilera2014} D. Aguilera {\it et al.}, Class. Quantum Grav. {\bf 31}, 115010 (2014) 

\bibitem{Delva2006} P. Delva, M.C. Angonin, P. Tourrenc, Phys. Lett. A {\bf 357}, 249 (2006)

\bibitem{Tino2007} G. M. Tino and F. Vetrano, Class. Quantum Grav. {\bf 24}, 2167 (2007)

\bibitem{Dimopoulos2008} S. Dimopoulos, P. W. Graham, J. M. Hogan, and M. A. Kasevich, S. Rajendran, Phys. Rev. D {\bf 78}, 122002 (2008)

\bibitem{Graham2013} P. W. Graham, J. M. Hogan, M. A. Kasevich, and S. Rajendran, Phys. Rev. Lett. {\bf 110}, 171102 (2013)

\bibitem{Foster2002} G. T. Foster, J. B. Fixler, J. M. McGuirk, and M. A. Kasevich, Optics Letters {\bf 27}, 951 (2002)

\bibitem{Stockton2007} J. K. Stockton, X. Wu and M. A. Kasevich, Phys. Rev. A {\bf 76}, 033613 (2007)

\bibitem{Varoquaux2009} G. Varoquaux, R.A. Nyman, R. Geiger, P. Cheinet, A. Landragin and P. Bouyer, New Journal of Physics {\bf 11}, 113010 (2009) 


\bibitem{Chen2014} X. Chen, J. Zhong, H. Song, L. Zhu, J. Wang, and M. Zhan, Phys. Rev. A {\bf 90}, 023609 (2014)

\bibitem{Rosi2015} G. Rosi, L. Cacciapuoti, F. Sorrentino, M. Menchetti, M. Prevedelli, and G.M. Tino, Phys. Rev. Lett. {\bf 114}, 013001 (2015)

\bibitem{Duan2014} X.-C. Duan, M.-K. Zhou, D.-K. Mao, H.-B. Yao, X.-B. Deng, J. Luo, and Z.-K. Hu, Phys. Rev. A {\bf 90}, 023617 (2014)

\bibitem{Biedermann2014} G. W. Biedermann, X. Wu, L. Deslauriers, S. Roy, C. Mahadeswaraswamy, M. A. Kasevich, Phys. Rev. A {\bf 91}, 033629 (2015)

\bibitem{LeGouet2008} J. Le Gou\"et, T. E. Mehlst\"aubler, J. Kim, S. Merlet, A. Clairon, A. Landragin and F. Pereira Dos Santos, Appl. Phys. B {\bf 92}, 133 (2008)

\bibitem{Geiger2011} R. Geiger, V. M{\'e}noret, G. Stern, N. Zahzam, P. Cheinet, B. Battelier, A. Villing, F. Moron, M. Lours, Y. Bidel, A. Landragin, P. Bouyer, Nature Comm. {\bf 2}, 474 (2011)

\bibitem{Lautier2014} J. Lautier, L. Volodimer, T. Hardin, S. Merlet, M. Lours, F. Pereira Dos Santos, and A. Landragin, Appl. Phys. Lett. {\bf 105}, 144102 (2014)

\bibitem{Sorrentino2012} F. Sorrentino, A. Bertoldi, Q. Bodart, L. Cacciapuoti, M. d. Angelis, Y. Lien, M. Prevedelli, G. Rosi and G. M. Tino, Appl. Phys. Lett. {\bf 101}, 114106 (2012)

\bibitem{Merlet2009} S. Merlet, J. Le Gou{\"e}t, Q. Bodart, A. Clairon, A. Landragin, F. Pereira dos Santos, P. Rouchon, Metrologia {\bf 46}, 87 (2009)

\bibitem{Chiow2009} S. W. Chiow, S. Herrmann, S. Chu, and H. M{\"u}ller, Phys. Rev. Lett. {\bf 103}, 050402 (2009)

\bibitem{Barrett2015} B. Barrett, L. Antoni-Micollier, L. Chichet, B. Battelier, P.-A. Gominet, A. Bertoldi, P. Bouyer, and A. Landragin, ArXiv:1503.08423


\end{thebibliography}
\end{document}